\documentclass[twocolumn,aps,prb,groupedaddress,superscriptaddress,amsmath,floatfix,amssymb,showpacs,noeprint,english]{revtex4-2}
\usepackage{graphicx}
\usepackage{dcolumn}
\usepackage{bm}
\usepackage[T1]{fontenc}
\usepackage{mathrsfs}
\usepackage{amsbsy}
\usepackage{amstext}
\usepackage{amsmath, amsthm, amsfonts, amssymb, bbold}
\usepackage{setspace}
\usepackage{esint}
\usepackage{xcolor,colortbl}
\usepackage{float}
\usepackage{placeins}
\usepackage[colorlinks,citecolor=blue,urlcolor=blue,linkcolor=blue,hypertexnames=true]{hyperref} 
\usepackage{babel}
\usepackage{booktabs}
\usepackage{tikz}

\usepackage{tikz}
\usepackage{quantikz}

\newcommand{\tbeta}[0]{\tilde{\beta}}
\newcommand{\tH}[0]{\mathcal{H}}
\newcommand{\tU}[0]{\mathcal{U}}
\newcommand{\df}[0]{\dot{f}}

\begin{document}

\title{Accelerating Feedback-Based Quantum Algorithms through Time Rescaling}

\author{L. A. M. Rattighieri}
\altaffiliation{These authors contributed equally to this work}
\affiliation{Faculty of Sciences, UNESP - S{\~a}o Paulo State University, 17033-360 Bauru-SP, Brazil}

\author{G. E. L. Pexe}
\altaffiliation{These authors contributed equally to this work}
\affiliation{Faculty of Sciences, UNESP - S{\~a}o Paulo State University, 17033-360 Bauru-SP, Brazil}

\author{B. L. Bernado}
\affiliation{Department of Physics, Federal University of Paraíba, 58051-900 João Pessoa-PB, Brazil}

\author{F. F. Fanchini}
\email{felipe.fanchini@unesp.br}
\affiliation{Hospital Israelita Albert Einstein, 05652-900 São Paulo-SP, Brazil}
\affiliation{Faculty of Sciences, UNESP - S{\~a}o Paulo State University, 17033-360 Bauru-SP, Brazil}
\affiliation{QuaTI - Quantum Technology \& Information, 13560-161 São Carlos-SP, Brazil}

\date{\today}

\begin{abstract}
This work investigates the impact of time rescaling on the performance of Feedback Quantum Algorithms (FQA) and their variant for optimization tasks, FALQON. We introduce TR-FQA and TR-FALQON, time-rescaled versions of FQA and FALQON, respectively. The method is applied to two representative problems: the MaxCut combinatorial optimization problem and ground-state preparation in the ANNNI quantum many-body model. The results show that TR-FALQON accelerates convergence to the optimal solution in the early layers of the circuit, significantly outperforming its standard counterpart in shallow-depth regimes. In the context of state preparation, TR-FQA demonstrates superior convergence, reducing the required circuit depth by several hundred layers. These findings highlight the potential of time rescaling as a strategy to enhance algorithmic performance on near-term quantum devices.
\end{abstract}

\maketitle

\section{Introduction}

Quantum algorithms have demonstrated significant potential in solving optimization problems more efficiently than classical methods. However, their practical implementation still faces major challenges, particularly due to the increasing circuit depth required to execute the desired operations \cite{RevModPhys.94.015004, Cerezo2021}. This depth makes quantum algorithms more susceptible to errors, a critical limitation for noisy intermediate-scale quantum (NISQ) devices, which remain constrained by limited coherence times and imperfect gate fidelity.

Despite the numerous potential applications of quantum computing, one of the most promising areas, capable of impacting a wide range of industries, is quantum optimization. Traditional approaches such as the Variational Quantum Eigensolver \cite{Peruzzo2014, PhysRevA.104.062426, PhysRevA.110.022430} and the Quantum Approximate Optimization Algorithm \cite{qaoa, ichikawa2025optimalelementalconfigurationsearch, 10.1145/3603273.3631193} have become widely used in this context. These algorithms rely on a hybrid quantum-classical loop, where a classical optimizer is used to update the parameters of a parameterized quantum circuit. In contrast, feedback-based quantum algorithms, such as the Feedback Quantum Algorithm (FQA) \cite{larsen2023feedbackbased, PhysRevB.110.224422} and its specialization for optimization tasks, the Feedback-Based Algorithm for Quantum Optimization (FALQON) \cite{1Magann_2022}, offer an alternative strategy by eliminating the need for a classical optimization routine. Instead, they iteratively adjust circuit parameters based on measurements, guiding the system toward the desired state through a closed-loop quantum control mechanism.

While feedback-based algorithms offer a promising alternative to hybrid quantum-classical methods, they are not without limitations. A key challenge lies in the circuit depth required for their implementation: in general, a large number of layers is necessary to achieve satisfactory results. This limits their applicability on quantum devices that suffer from noise and finite coherence times. To overcome this drawback, several strategies have been proposed. For instance, the FOCQS method \cite{brady2024focqsfeedbackoptimallycontrolled} applies a perturbative update of control layers based on Pontryagin’s optimal control, aiming to reduce circuit depth and improve convergence. Another approach \cite{PhysRevResearch.7.013035} modifies the feedback rule by incorporating a second-order time approximation, which enables larger time steps and enhances convergence, leading to a reduced circuit depth in problems such as MaxCut. Furthermore, counterdiabatic control \cite{PhysRevResearch.6.043068} has been investigated as a way to accelerate system evolution, potentially reducing both execution time and circuit depth.

In this work, we pursue a similar direction and propose a modification of the system’s time evolution aimed at accelerating dynamics and reducing the number of required layers. Our approach is based on shortcuts to adiabaticity (STA), a class of techniques widely used to optimize the evolution of quantum systems \cite{rt13, rt14, rt15}, with successful applications across diverse platforms, including cold atoms \cite{rt16}, trapped ions \cite{rt17}, nitrogen-vacancy centers \cite{rt7}, and superconducting qubits \cite{rt18}. Building on this framework, we introduce the Time-Rescaled Feedback Quantum Algorithm (TR-FQA), which enhances the efficiency of the standard FQA by modifying the time dependence of the reference Hamiltonian, effectively implementing an STA protocol that improves convergence and reduces circuit depth. When applied to optimization problems, this method is referred to as TR-FALQON. To illustrate its effectiveness, we test TR-FALQON on the MaxCut problem and TR-FQA on the ANNNI model, demonstrating improvements in convergence and computational efficiency.

This manuscript is organized as follows. In Sec. II, we provide a brief overview of Quantum Lyapunov Control and its connection to FQA. In Sec. III, we introduce the Time Rescaling Method and present the formulation of TR-FQA.  In Sec. IV, we describe two case studies used to evaluate our method: the MaxCut problem, representing a combinatorial optimization task, and the ANNNI model, which involves quantum ground-state preparation.  In Sec. V, we present our numerical results, highlighting the improvements brought by TR-FALQON and TR-FQA. Finally, in Sec. VI, we summarize our findings and discuss potential future research directions.

\section{Quantum Lyapunov Control and Feedback Quantum Algorithm}

\begin{figure*}[htpb]
    \centering
    \includegraphics[width=1\linewidth]{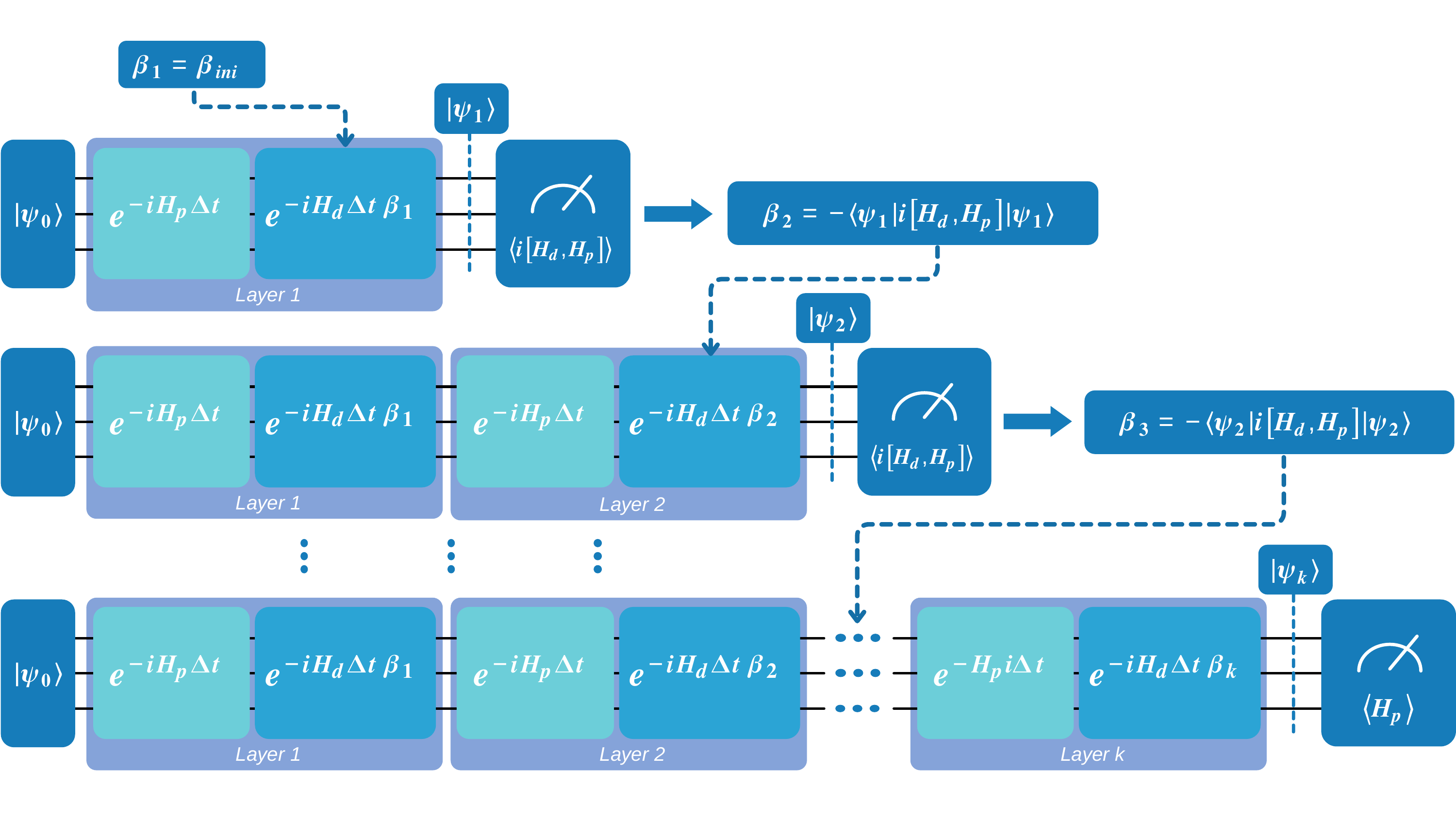}
    \caption{Illustrative diagram of the FQA \cite{PhysRevB.110.224422}. The process begins with the state $\ket{\psi_0}$, and at each layer $k$, the unitary operators $e^{-iH_p \Delta t}$ and $e^{-iH_d \Delta t \beta_k}$ are applied sequentially. The parameter $\beta_k$ is adaptively adjusted at each iteration. This dynamic is repeated iteratively, guiding the evolution of the state $\ket{\psi}$ through the layers until the desired solution is reached.}
    \label{fig:diag_falqon}
\end{figure*}

Quantum Lyapunov Control (QLC) \cite{inproceedings, PhysRevA.106.062414} is a method used to determine control functions that asymptotically guide the evolution of a quantum system toward a desired target state. To illustrate the principle of QLC, we consider a system governed by the time-dependent Hamiltonian $H(t) = H_p + \beta(t) H_d$, whose dynamics is governed by the time-dependent Schrödinger equation (with $\hbar = 1$):
\begin{equation}
    i \frac{d}{dt} \ket{\psi(t)} = (H_p + \beta(t) H_d) \ket{\psi(t)}.
    \label{eqn:equacao_sch}
\end{equation}
Here, $\ket{\psi(t)}$ is the state vector of the system at time $t$. The operators $H_p$ and $H_d$ are dimensionless Hamiltonians, where $H_p$ encodes the problem to be solved (the problem Hamiltonian), and $H_d$ represents the driving or control Hamiltonian. The scalar function $\beta(t)$ is a time-dependent control parameter that couples to $H_d$. The objective is to design $\beta(t)$ such that the system evolves toward the ground state of $H_p$ by minimizing an objective function $J(t)$, known as the Lyapunov function. In this context, $J(t)$ is defined as the expectation value of $H_p$:
\begin{equation}
    J(t) = \bra{\psi(t)} H_p \ket{\psi(t)}.
\end{equation}
By minimizing $J(t)$, the system is driven toward the ground state of $H_p$. The control function $\beta(t)$ is chosen to ensure that $J(t)$ decreases monotonically over time, i.e.,
\begin{equation}
    \frac{dJ(t)}{dt} \le 0, \hspace{0.3cm}\forall \,t.
    \label{eqn:condicao_lyapunov}
\end{equation}
Calculating the time derivative of $J(t)$ yields:
\begin{equation}
    \frac{dJ(t)}{dt} = A(t) \beta(t),
\end{equation}
where $A(t) = \bra{\psi(t)} i [H_d, H_p] \ket{\psi(t)}$.
To guarantee the condition in Eq.~\eqref{eqn:condicao_lyapunov}, we require:
\begin{equation}
    \beta(t) = - \omega f(t, A(t))
    \label{eqn:beta_geral}
\end{equation}
where $\omega > 0$ and $f(t, A(t))$ is a continuous function satisfying $f(t, 0) = 0$ and $A(t) f(t, A(t)) > 0$ for all $A(t) \neq 0$.  This ensures that $J(t)$ decreases monotonically, driving the system toward the target state.

Equation (\ref{eqn:beta_geral}) allows flexibility in the choice of $\beta(t)$. The simplest and most common option is to set $\omega = 1$ and choose $f(t, A(t)) = A(t)$, leading to:
\begin{equation}
    \beta(t) = - A(t)
    \label{eqn:beta_especifico}
\end{equation}
a choice that clearly satisfies the condition in Eq.~\eqref{eqn:condicao_lyapunov}, since $\frac{dJ(t)}{dt} = - [A(t)]^2 \leq 0$. It is important to emphasize that $H_p$ and $H_d$ must not commute, otherwise, $A(t) = 0$ for all $t$, and the system would no longer evolve toward the ground state, as the cost function $J(t)$ would remain constant.

Building on this control strategy, the FQA was developed as a quantum algorithm tailored for ground-state preparation. It is based on quantum Lyapunov control and iteratively constructs a parameterized quantum circuit. In contrast to variational quantum algorithms, which depend on a classical optimizer to adjust circuit parameters, the FQA adopts a fully quantum feedback-based mechanism. At each iteration, the parameters of a new layer of quantum gates are updated based on measurements of the state produced by the previous layers. This continuous feedback process steers the system toward the ground state of the problem Hamiltonian.  

The algorithm's dynamics can be understood from the general solution of the Schrödinger equation, Eq. (\ref{eqn:equacao_sch}):
\begin{equation}
    \ket{\psi(t)} = \mathcal{T} e^{\int^t_0 {-i(H_p + \beta(t')H_d) dt'}} \ket{\psi(0)},
\end{equation}
where $\mathcal{T}$ is the time-ordering operator. Discretizing the time interval $[0,t]$ into a sequence of $k$ steps of duration $\Delta t$ and applying a Trotter-Suzuki decomposition \cite{Schuld:2021mml}, we obtain the discretized form:
\begin{equation}
\ket{\psi_k} = U_d(\beta_k) U_p \dots U_d(\beta_2) U_p U_d(\beta_1) U_p \ket{\psi_0},
\label{eq_ev_trottenizada}
\end{equation}
where $U_d(\beta_k) = e^{-i \beta_k H_d \Delta t}$ and $U_p = e^{-i H_p \Delta t}$, with $\beta_k$ given by the control parameter associated with the $k$-th layer ($\beta_k \approx \beta(k\Delta t)$) and $\ket{\psi_k}$ given by the state of the system after the application of the $k$-th layer ($\ket{\psi_k} \approx \ket{\psi(k\Delta t)}$). To ensure that the system evolves toward the ground state of $H_p$, we define $\beta_k$ as:
\begin{equation}
    \beta_k = - A_{k-1} = - \bra{\psi_{k-1}} H_p \ket{\psi_{k-1}},
    \label{eqn:lei_feedback0}
\end{equation}
thus establishing the Feedback Law, which guides the construction of subsequent layers.  

The FQA begins with the preparation of the circuit in an initial state $\ket{\psi_0}$, which should be easy to generate. After selecting an appropriate time interval $\Delta t$ and defining the number of layers $\ell$, the first layer is applied, consisting of the unitary operators $e^{-iH_p \Delta t}$ and $e^{-iH_d \Delta t}$, where $\beta_1 = 0$. For each new layer $k$, the operators $e^{-iH_p \Delta t}$ and $e^{-iH_d \Delta t \beta_k}$ are applied, with $\beta_k$ adjusted according to Eq. (\ref{eqn:lei_feedback0}), using the state generated by the previous layer. The term ``feedback'' refers to this process, in which the parameters of subsequent layers are determined based on measurements of the current state. This cycle is repeated iteratively until the state approaches the ground state, with the cost function $J_k = \left\langle\psi_k\left|H_p\right|\psi_k\right\rangle$ decreasing at each iteration. Figure \ref{fig:diag_falqon} illustrates the operation of the FQA.

Having detailed the structure and operation of the standard FQA, we now turn to the improvements introduced in this work. In the following section, we present the Time Rescaling Method, which serves as the foundation for enhancing the algorithm’s efficiency and reducing circuit depth.

\section{Time Rescaling Method}

\begin{figure*}
    \centering
    \includegraphics[width=1\linewidth]{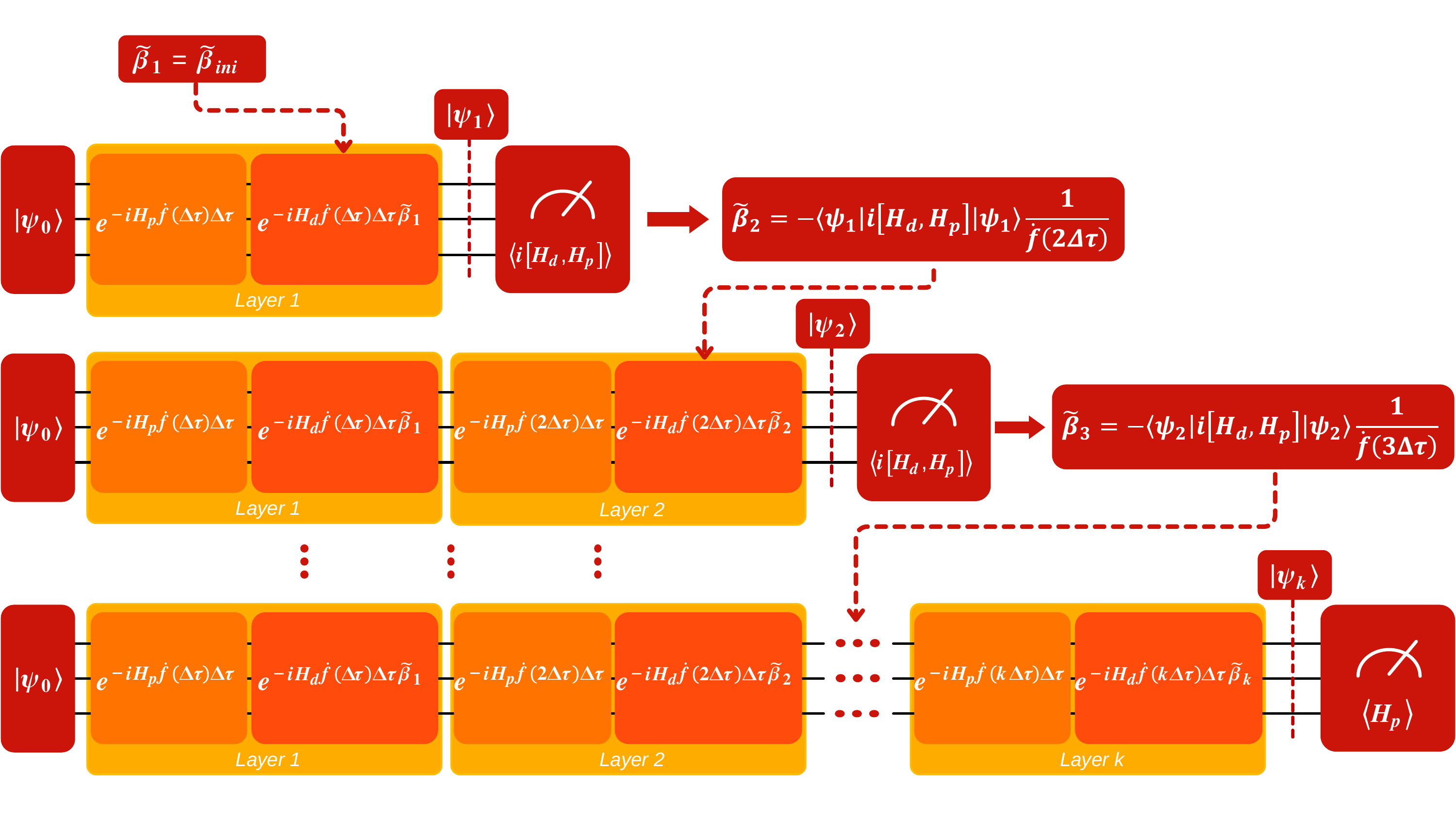}
    \caption{Illustrative diagram of the TR-FQA. The process starts with the initial state $\ket{\psi_0}$. In each layer $k$, the unitary operators $e^{-iH_p \dot{f}(k\Delta \tau) \Delta \tau}$ and $e^{-iH_d \dot{f}(k\Delta \tau) \Delta \tau \tilde{\beta}_k}$ are applied sequentially, adaptively adjusting the parameter $\tilde{\beta}_k$.}
    \label{fig:diagrama_tr-FALQON}
\end{figure*}

The Time Rescaling Method was introduced by Bertulio in \cite{PhysRevResearch.2.013133} and further developed by Ferreira in \cite{ferreira2024shortcutsadiabaticitydesignedtimerescaling}. This method modifies the time dependence of a quantum system's Hamiltonian, altering its evolution dynamics and allowing the target final state to be reached more efficiently, either by accelerating or decelerating the process as needed.

To formalize this idea, consider a quantum system governed by a time-dependent Hamiltonian $H(t)$, evolving within the time interval $t \in [0, t_f]$ and described by the Schrödinger equation. The corresponding unitary evolution is given by:
\begin{equation}
    U\left(t_f\right) = \mathcal{T} \exp \left\{-\frac{i}{\hbar} \int_0^{t_f} H\left(t^{\prime}\right) d t^{\prime}\right\}.
\end{equation}  
The system’s dynamics can be modified through a rescaling function $t = f(\tau)$, which redefines the temporal evolution. Applying this change of variables, the unitary evolution can be rewritten as:
\begin{equation}
    \mathcal{U}\left(t_f, 0\right) = \mathcal{T} \exp \left\{-\frac{i}{\hbar} \int_{f^{-1}(0)}^{f^{-1}\left(t_f\right)} H(f(\tau)) \df(\tau) d \tau\right\}.
    \label{eqn:evolucao_reescalonada}
\end{equation}  
This expression describes the evolution of a system governed by the rescaled Hamiltonian:
\begin{equation}
    \tH(\tau) = H(f(\tau)) \df(\tau),
\end{equation}  
which depends on the rescaling function $f(\tau)$ and its time derivative $\dot{f}(\tau)$. Thus, if the evolution of the rescaled system occurs in the interval $\tau \in [f^{-1}(0), f^{-1}(t_f)]$, it will produce the same final state $\ket{\psi(t_f)}$ as the original evolution of $H(t)$, provided that both start from the same initial state $\ket{\psi(0)}$.

This implies that, by appropriately choosing $f(\tau)$,  the process governed by $\tH(\tau)$ can be accelerated ($\Delta \tau < \Delta t$) or decelerated ($\Delta \tau > \Delta t$) relative to the original evolution, where $\Delta t = t_f - 0$. If the original evolution corresponds to an adiabatic transformation, and the function $f(\tau)$ is chosen such that: (i) the initial times coincide, i.e., $f^{-1}(0) = 0$; (ii) the final time is shortened, $f^{-1}(t_f) < t_f$; and (iii) the initial and final Hamiltonians remain unchanged, $\tH(0) = H(0)$ and $\tH(f^{-1}(t_f)) = H(t_f)$, then the time rescaling method can be classified as a shortcut to adiabaticity.

A common choice for the rescaling function is:
\begin{equation}
f(\tau) = a \tau - \frac{t_f}{2\pi a}(a - 1) \sin \left( \frac{2 \pi a}{t_f} \tau \right),
\label{f1}
\end{equation}  
where $a$ is a parameter that controls the temporal contraction. Another possibility is a polynomial function: 
\begin{equation}
f(\tau) = \frac{2(a^2 - a^3)}{t_f^2} \tau^3 + \frac{3(a^2 - a)}{t_f} \tau^2 + \tau.
\label{f2}
\end{equation}  
These functions allow modifying the system's temporal evolution without altering its final state, with the total evolution time given by $\Delta \tau = \Delta t / a$. 

Now that we have described the core idea behind the time rescaling method, we proceed to adapt it within the FQA framework. By modifying the algorithm's evolution dynamics, this adaptation enhances performance and allows the desired solution to be reached more efficiently, reducing the computational time.

\subsection{Time-Rescaled Feedback Quantum Algorithm}

Based on the time rescaling method, we can adapt it for use with FQA. For this, we consider the Schrödinger equation applied to a time-dependent Hamiltonian in the form $H(t) = H_p + \beta(t) H_d$, where $H_p$ represents the problem Hamiltonian, $H_d$ is the driving Hamiltonian, and $\beta(t)$ is the control function. The evolution equation is given by:
\begin{equation}
    \frac{d}{dt} \ket{\psi(t)} = -i H(t) \ket{\psi(t)}.
\end{equation}
By rescaling the time through the transformation $t = f(\tau)$, the equation above becomes:
\begin{equation}
\begin{aligned}
    \frac{d}{d\tau} \ket{\psi(\tau)} &= -i \tH(\tau) \ket{\psi(\tau)}, \\
    &= -i H(f(\tau)) \dot{f}(\tau) \ket{\psi(\tau)}.
\end{aligned}
\end{equation}

To obtain the control function for this case, we follow the same steps used in the FQA. First, we calculate the time derivative of the expectation value of $H_p$ along the rescaled evolution:
\begin{equation}
\begin{aligned}
    \frac{d}{d\tau} \bra{\psi(\tau)} H_p \ket{\psi(\tau)} &= i \bra{\psi(\tau)} \tH(\tau) H_p - H_p \tH(\tau) \ket{\psi(\tau)}, \\
    &= i \bra{\psi(\tau)} [\tH(\tau), H_p] \ket{\psi(\tau)}, \\
    &= i \dot{f}(\tau) \bra{\psi(\tau)} [H(f(\tau)), H_p] \ket{\psi(\tau)}, \\
    &= \beta(f(\tau)) \dot{f}(\tau) \bra{\psi(\tau)} i [H_d, H_p] \ket{\psi(\tau)}, \\
    &= \tbeta(\tau) A(\tau) \dot{f}(\tau),
\end{aligned}
\end{equation}

\noindent
where we define $A(\tau) = \langle \psi(\tau) | i [H_d, H_p] | \psi(\tau) \rangle$, which quantifies the non-commutativity between $H_d$ and $H_p$, and $\tbeta(\tau) = \beta(f(\tau))$, representing the control function evaluated at the rescaled time.

To ensure that the cost function, defined as the expectation value of $H_p$, decreases throughout the evolution, we determine a control function $\tbeta(\tau)$ that satisfies:
\begin{equation}
    \frac{d}{d\tau} \bra{\psi(\tau)} H_p \ket{\psi(\tau)} \leq 0, \quad \forall \tau.
\end{equation}
A choice that satisfies this condition is:
\begin{equation}
    \tbeta(\tau) = -\omega \, F(\tau, A(\tau)) \, G(\dot{f}(\tau)),
\end{equation}
where $\omega > 0$, $F(\tau, A(\tau))$ is a continuous function with $F(\tau, 0) = 0$ and $A(\tau)F(\tau, A(\tau)) \ge 0$ for all $A(\tau) \neq 0$, and $G(\dot{f}(\tau))$ is a continuous function with $G(\dot{f}(\tau)) \dot{f}(\tau) \ge 0$. We will adopt $\omega = 1$, $F(\tau, A(\tau)) = A(\tau)$, and $G(\dot{f}(\tau)) = 1 / \dot{f}(\tau)$, so that $\tbeta(\tau)$ becomes:
\begin{equation}
    \tbeta(\tau) = - A(\tau) \frac{1}{\dot{f}(\tau)}.
\end{equation}
The choice of $\tbeta(\tau)$ in this way ensures that the function $J(\tau)$ decreases over time.

Now, we can adapt the FQA circuit to incorporate the rescaled time. For this, we consider the rescaled time evolution in the interval $\tau_0 = 0$ to $\tau_f = f^{-1}(t_f)$, discretized into $k$ steps with time interval $\Delta \tau$. This evolution can be described by a sequence of applications of the unitary operator:
\begin{equation}
    \tU(\tau) = \exp{\left(-i(H_p + \tbeta(\tau)H_d) \df (\tau)\Delta \tau \right)}.
\end{equation}
The application of $\tU(\tau)$ on the state $\ket{\psi(\tau)}$ produces the updated state $\ket{\psi(\tau + \Delta \tau)}$. To implement $\tU(\tau)$ in a quantum circuit, we use a Trotterized approximation, which allows us to express $\tU(\tau)$ as:
\begin{equation}
    \tU(\tau) = \tU_d (\tbeta(\tau)) \tU_p
\end{equation}
where $\tU_d (\tbeta(\tau)) = \exp{\left(-iH_d \tbeta(\tau)\df (\tau)\Delta \tau \right)}$ and $\tU_p = \exp{\left(-iH_p \df (\tau)\Delta \tau \right)}$. From this decomposition, we can describe the resulting state of the FQA circuit after a sequence of $k$ applications of these Trotterized operators:
\begin{equation}
    \ket{\psi_k} = \tU_d (\tbeta_k) \tU_p \;\; \dots \;\; \tU_d (\tbeta_2) \tU_p \; \tU_d (\tbeta_1) \tU_p \ket{\psi_0},
\end{equation}
where $\ket{\psi_k}$ is the state of the circuit after the $k$-th layer application ($\ket{\psi_k} \approx \ket{\psi(k\Delta \tau)}$) and $\beta_k$ is the control parameter for the $k$-th layer ($\tbeta_k \approx  \tbeta(k\Delta \tau)$). Also, similarly to the standard FQA, since $\beta_k$ is required in advance to construct the state $\ket{\psi_k}$, we use the state from the previous iteration. For sufficiently small time intervals, this state closely approximates $\ket{\psi_k}$, allowing us to define the feedback law as follows:
\begin{equation}
    \tbeta_k = - A_{k-1} \, \frac{1}{\df(k\Delta \tau)}  = - \bra{\psi_{k-1}} H_p \ket{\psi_{k-1}} \frac{1}{\df(k\Delta \tau)}.
    \label{eqn:lei_feedback}
\end{equation}

In this way, we are able to adapt the FQA to incorporate time re-scaling. Figure \ref{fig:diagrama_tr-FALQON} illustrates the operation of the TR-FQA.

As we can see, the TR-FQA leverages time rescaling to accelerate convergence, improving the efficiency of FQA by reducing the total evolution time. Moreover, since the rescaling function depends on both $a$ and $t_f$, the behavior of the TR-FQA also varies with these parameters, which must be carefully tuned to achieve the desired performance.

\section{Applications to Optimization and State Preparation}

To illustrate the effectiveness and versatility of the time-rescaled feedback algorithm, we consider two distinct problems. The first is a well-known combinatorial optimization problem, MaxCut, which serves as a benchmark for evaluating the performance of TR-FALQON in the context of optimization. The second is the ANNNI model, a quantum many-body system used to test the ability of TR-FQA to prepare ground states.

\subsection{MaxCut}  

The MaxCut problem \cite{Bechtold_2023, lucas2014} is a combinatorial optimization problem on graphs. Given a graph $G = (V, E)$, the goal is to partition the set of vertices $V$ into two subsets, $S$ and $S'$, such that the number of edges $e \in E$ connecting vertices in different subsets is maximized. If the edges have weights, the objective is to maximize the sum of the weights of the edges crossing the partition, known as the cut.  

To solve MaxCut using quantum algorithms, the problem is mapped onto a quantum Hamiltonian. Each vertex $i \in V$ is associated with a qubit, where the states $|0\rangle$ and $|1\rangle$ represent the two subsets of the partition: $|0\rangle$ means the vertex belongs to one subset, while $|1\rangle$ means it belongs to the other. The problem can then be formulated as an Ising Hamiltonian:  
\begin{equation}
    H = -\sum_{i,j = 0}^{L-1} \frac{w_{ij}}{2} (\mathbb{1} - \sigma_z^i \sigma_z^j).
    \label{eqn:H_maxcut}
\end{equation}  
Here, $w_{ij}$ represents the weight of the edge connecting vertices $i$ and $j$. If the graph is unweighted, $w_{ij} = 1$ for all edges. The term $(\mathbb{1} - \sigma_z^i \sigma_z^j)$ evaluates to 1 when the qubits are in different states ($|0\rangle$ and $|1\rangle$), meaning the corresponding edge is cut, and 0 when they are in the same state, meaning the edge is not cut. The ground state of this Hamiltonian corresponds to the optimal solution of MaxCut.

\subsection{ANNNI Model}

The Axial Next-Nearest Neighbor Ising, known as the ANNNI model \cite{SELKE1988213, Suzuki2013}, is an extension of the traditional Ising model. It describes the behavior of spins-1/2 arranged in a one-dimensional chain, where the spins interact anisotropically with both their nearest and next-nearest neighbors.

The Hamiltonian that describes the ANNNI model is given by:
\begin{equation}
    H_{A}(\kappa, g) = -J \sum_{j=0}^{L-1} \left( \sigma_j^z \sigma_{j+1}^z - \kappa \sigma_j^y \sigma_{j+2}^y + g \sigma_{j}^x \right)
    \label{eqn:H_annni}
\end{equation}
where $L$ is the size of the spin chain and $\sigma_j^x$, $\sigma_j^y$, and $\sigma_j^z$ are the Pauli matrices corresponding to the spin components at site $j$. In this work, we adopt periodic boundary conditions, such that $\sigma_{L} = \sigma_{0}$ and $\sigma_{L+1} = \sigma_1$. The parameter $J$ is the coupling constant for the ferromagnetic nearest-neighbor interaction, which is positive and defines the energy scale ($J = 1$). The term $\kappa$ is the dimensionless coupling constant that describes the strength of the antiferromagnetic interaction between the next-nearest neighbors. The term $g$ is a dimensionless coupling constant that describes the strength of the transverse magnetic field acting on the system.

\section{Results}

In this section, we present the analyses performed with TR-FQA, dividing them into two parts. In the first part, we investigate the algorithm's performance in solving optimization problems, referring to it as TR-FALQON. As a test case, we consider the MaxCut problem, using the Hamiltonian defined in Eq. (\ref{eqn:H_maxcut}) as $H_p$. In the second part, we evaluate the effectiveness of TR-FQA in quantum state preparation, analyzing its performance in the ANNNI model, where we adopt the Hamiltonian described in Eq. (\ref{eqn:H_annni}) as $H_p$. In both simulations, the driving Hamiltonian used was:
\begin{equation}
H_d = \sum^{L}_{j=1} \sigma_j^x.
\end{equation}

The system was initially prepared in the state:
\begin{equation}
\ket{\psi_0} = \prod_{i=1}^{L} \frac{1}{\sqrt{2}} (\ket{0} + \ket{1}),
\end{equation}
where $L$ corresponds to the number of qubits required for encoding the problem. In the case of MaxCut, $L$ is equal to the number of vertices in the graph, while for the ANNNI model, it corresponds to the number of sites in the chain.

\subsection{Application to the MaxCut Problem}

\begin{figure}
    \centering
    \includegraphics[width=1\linewidth]{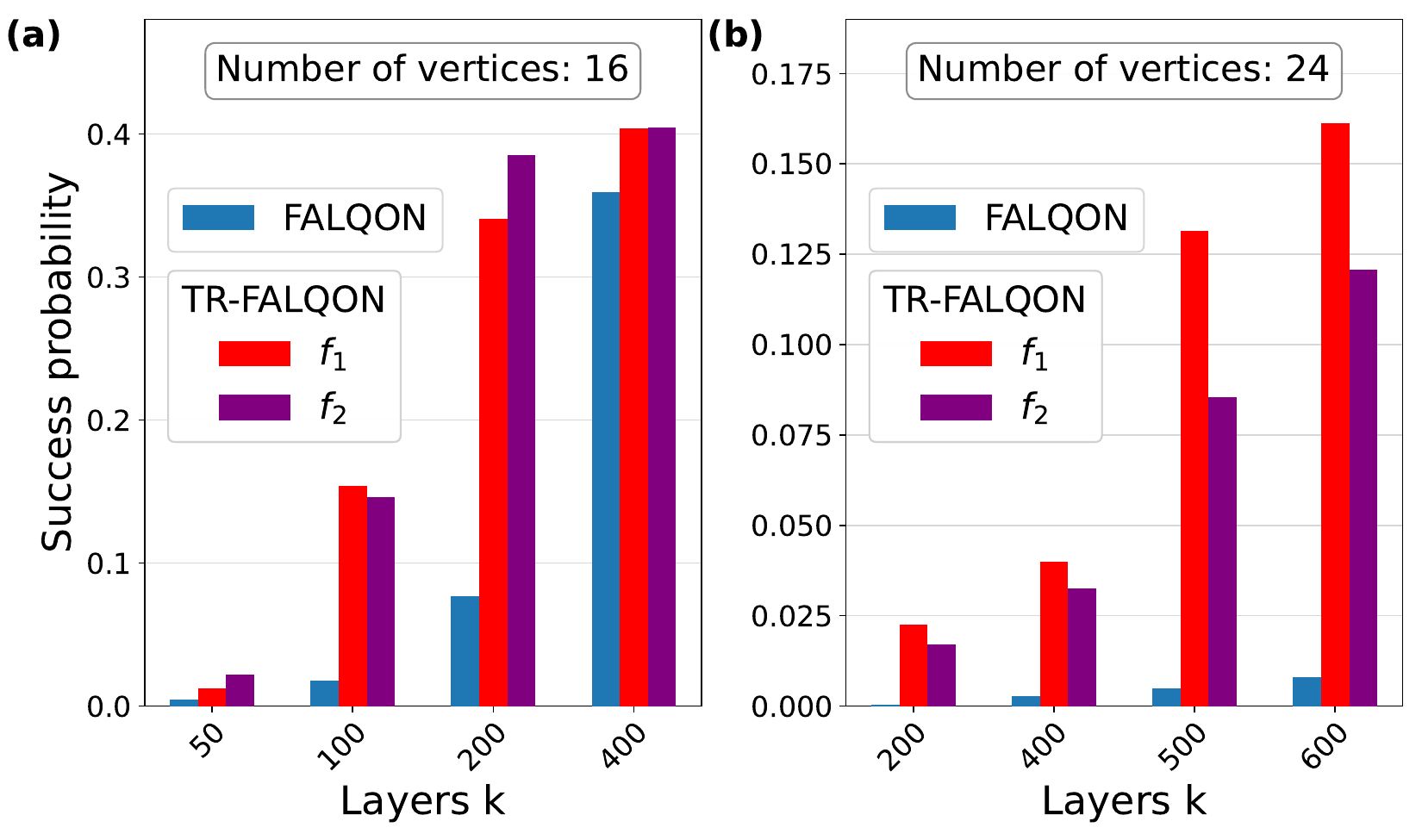}
    \caption{Comparison of the performance of FALQON and TR-FALQON for different time rescaling functions, showing the probability of obtaining the solution to the problem from the prepared state for different layer depths. Panel (a) presents the results for a 16-vertex graph, where FALQON was executed with $\Delta t = 0.04$, 400 layers, and TR-FALQON was executed with $\Delta \tau = 0.04$, 400 layers, $a = 2$, and $t_f = 16$. Panel (b) presents the results for a 24-vertex graph, where FALQON was executed with $\Delta t = 0.03$, 600 layers, and TR-FALQON was executed with $\Delta \tau = 0.03$, 600 layers, $a = 2$, and $t_f = 18$.}
    \label{fig:grafico1-tr}
\end{figure}

Figure \ref{fig:grafico1-tr} compares the performance of FALQON and TR-FALQON using the time rescaling functions $f_1$ (Eq.~\ref{f1}) and $f_2$ (Eq.~\ref{f2}). Panel \ref{fig:grafico1-tr}(a) shows the result for a 16-vertex graph. For shallow circuits (e.g., 50 layers), the probability of obtaining the correct solution is low across all approaches.  However, TR-FALQON already outperforms FALQON at this early stage. As the number of layers increases, the probability of success grows significantly for TR-FALQON, especially at 100 and 200 layers, where it consistently surpasses FALQON. At 400 layers, as expected, the advantage of TR-FALQON diminishes, and the performance of standard FALQON catches up. This occurs because, with sufficiently deep circuits, the system has enough evolution time to reach the ground state even without time rescaling, reducing the relative benefit of the shortcut-based approach.

Panel \ref{fig:grafico1-tr}(b) presents the results for a 24-vertex graph. In this case, TR-FALQON maintains a clear advantage throughout, with the success probability increasing more rapidly compared to FALQON as the number of layers grows. The rescaling function $f_1$ yields the best overall performance, demonstrating the potential of time rescaling to accelerate convergence in combinatorial optimization problems. Notably, in this case, the standard FALQON has not yet reached the performance of either rescaled method at this depth, indicating that the algorithm has not fully converged for the given number of layers.

\subsection{Application to the ANNNI Model}

\begin{figure}[!h]
    \centering
    \includegraphics[width=1\linewidth]{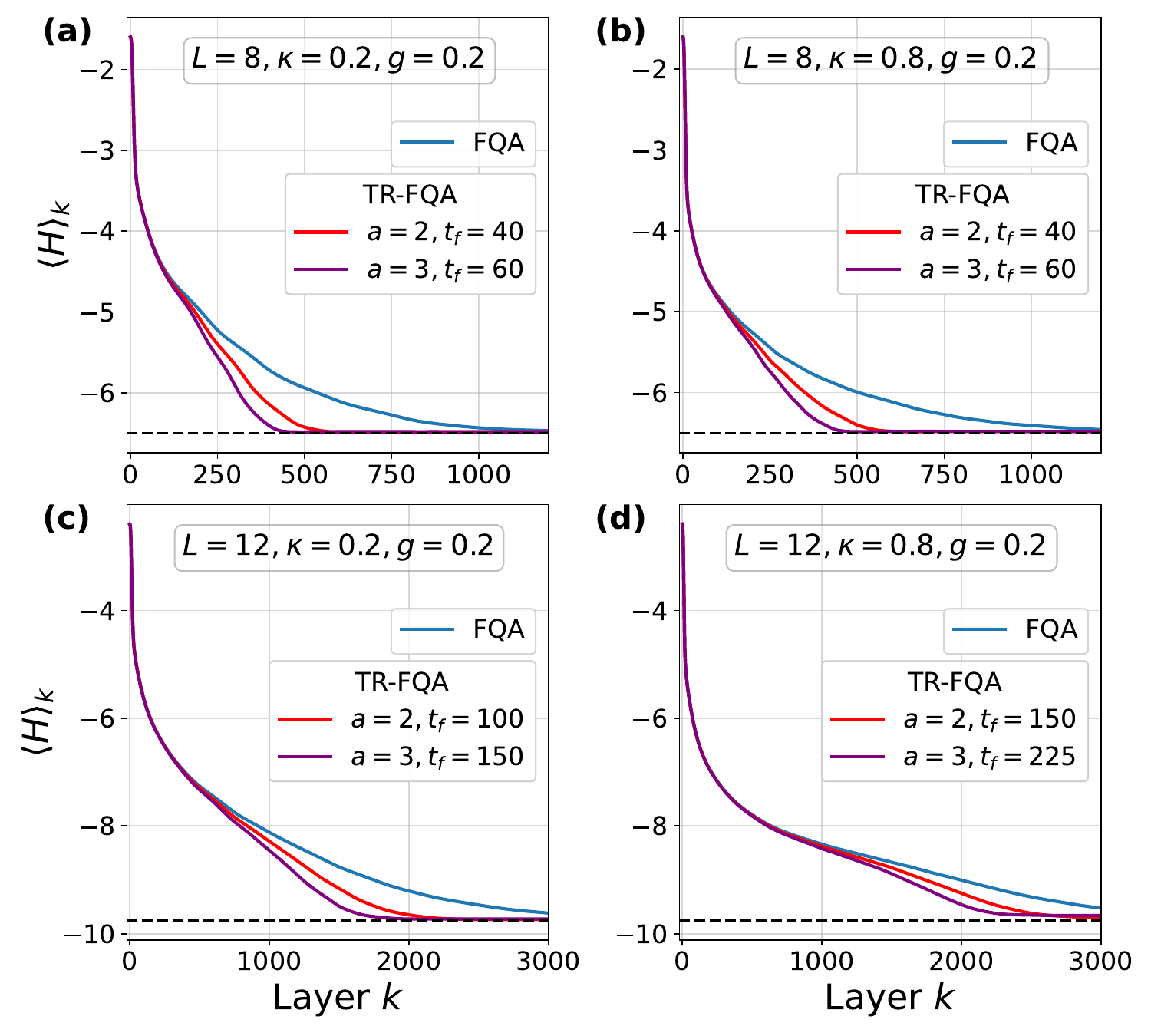}
    \caption{Numerical simulation comparing FQA and TR-FQA applied to the ANNNI model for different chain configurations and values of $\kappa$ and $g$. The TR-FQA simulations were performed using the function $f_1$, with the values of $a$ and $t_f$ specified in each panel. For chains with $L = 8$ sites, $\Delta t = \Delta \tau = 0.01$ was adopted, while for $L = 12$ sites, $\Delta t = \Delta \tau = 0.005$ was used. In the panels, the dashed line represents the ground state energy, while the curves show the convergence of the cost function $J = \langle\Psi_{k}|H|\Psi_{k}\rangle$ as a function of the number of layers $k$.}
    \label{fig:grafico2-tr}
\end{figure}

Figure~\ref{fig:grafico2-tr} explores the scalability of the FQA and TR-FQA algorithms when applied to the ANNNI model, using the time rescaling function $f_1$ and different chain sizes $L$. The simulations were performed for various configurations of $\kappa$ and $g$, aiming to evaluate the ability of each algorithm to drive the system toward its ground state.  The panels indicate that as the system size increases, a greater number of layers $k$ is required to achieve convergence. However, TR-FQA maintains superior performance compared to FQA, achieving faster convergence to the ground state energy. Among the rescaled protocols, the configuration with $a = 3$ is particularly effective, reducing the required circuit depth by more than 500 layers compared to the standard FQA. Moreover, the results highlight the importance of selecting an appropriate final time $t_f$, which must be increased for larger systems and higher values of $a$ in order to maintain efficient convergence.

\FloatBarrier

\section{Conclusion}  
In this work, we explored the impact of time rescaling on the performance of feedback-based quantum algorithms, introducing the TR-FQA and TR-FALQON variants as time-rescaled versions of FQA and FALQON. These algorithms were applied to two distinct tasks: solving a combinatorial optimization problem (MaxCut) and preparing ground states in quantum many-body systems (ANNNI model). Our results demonstrate that TR-FALQON significantly enhances convergence in the early stages of the circuit, achieving higher success probabilities with fewer layers. Also, in the context of ground-state preparation, TR-FQA consistently outperformed standard FQA. Notably, the time-rescaled approach was able to reduce circuit depth by more than 500 layers while maintaining accuracy. This highlights the potential of time rescaling, as the standard and rescaled methods yield comparable results only when the algorithms are already near convergence, a regime that demands substantially deeper circuits. This makes the proposed approach particularly valuable for near-term quantum devices, where circuit depth is a critical constraint.

\begin{acknowledgments}
F.F.F acknowledge support from Funda\c{c}{\~a}o de Amparo {\`a} Pesquisa do Estado de S{\~a}o Paulo (FAPESP), project number 2023/04987-6 and from ONR, project number N62909-24-1-2012. G. E. L. P. and L. A. M. R. acknowledge support from Coordenação de Aperfeiçoamento de Pessoal de Nível Superior (CAPES), Projects No. 88887.829788/2023-00 and No. 88887.143168/2025-00, respectively.
\end{acknowledgments}

\appendix

\bibliography{main}

\end{document}